# An S-matrix Formalism for the Nonclassical Optical Response of Plasmonic Nanowires

Xin Zheng, Christos Tserkezis, Christos Mystilidis, *Member*, *IEEE*, Guy A. E. Vandenbosch, *Fellow*, *IEEE*, Xuezhi Zheng, *Senior Member*, *IEEE*

***Abstract*— This paper reports a computational scheme which predicts the light-matter interaction in multiple mesoscopic plasmonic nanowires (NWs), each of which includes concentric or eccentric constituent wires. Lying at the core of the proposed scheme is the S-Matrix of each individual cylindrical interface. The optical response of the materials inside and outside each interface can be described by three popular mesoscopic models: the Hydrodynamic Drude model (HDM) and its diffusive variant, namely the Generalized Nonlocal Optical Response (GNOR) model, as well as the Surface Response Model (SRM). To take into account the relative displacements between the cylindrical interfaces, the addition theorem is applied to construct the main equation behind the algorithm, where the possibility of longitudinal waves intrinsic in the HDM and GNOR models is also included. The optical responses calculated by the proposed S-Matrix Method are compared with an in-house developed boundary element method (BEM) toolbox for a single gold NW and an eccentric dual-material NW, and excellent agreements are seen. The algorithm is further physically checked for a sodium nanolens trimer featuring large field focusing, and the effects of the HDM and the SRM are compared.**

## I. INTRODUCTION

With the advancement of nanofabrication technology, nanostructures have entered the deep (even the sub-) nanometric regime [1], [2], [3], which is characterized by geometric sizes of only a few nanometers [4], [5] or inter-element separations on the order of sub-1nm scale [6], [7]. This advancement has paved the way for numerous groundbreaking applications, including single-molecule-level chemistry [8], [9], surface-enhanced Raman scattering (SERS) [10], [11], enhanced light harvesting [12], [13], and higher-harmonic generation [14]. Typical structures include dimers [15], [16], the nanoparticle-on-mirror structure [8], [10], [11], plasmonic nanowires [17], [18] and nano lenses [19], [20], [21]. In the deep- and sub-nm regime, the classical material model (Local Response Approximation, LRA), without taking semi-classical effects such as nonlocality [16], [22], electron spill-in/out [23], [24] and Landau damping [15], [25] into consideration, prove inadequate in reproducing experimental spectra [26]. Taking the nonlocal response as an example, one prominent manifestation is the resonant excitation of longitudinal bulk plasmon modes in thin metallic films [27] and nanowires [28], which occurs above the plasma frequency $\omega_p$. Also, the observed frequency shifts [29] and spectrum broadening [30], [31] demonstrate urgent demands for new physical models to be implemented in numerical and analytical simulations.

The Time-Dependent Density Functional Theory (TDDFT) [32] can, in principle, be used to deal with these challenges within a full quantum mechanical framework. However, its application is computationally intensive and is therefore often restricted to the electrostatic (quasi-static) limit, although retarded TDDFT approaches have also become feasible (e.g., an infinitely large solid in [33] where Bloch type boundary conditions may be used), which significantly limits its use in plasmonics to small atomic clusters [34], [35] or highly symmetric systems whose size is on the ~1 nm scale [36], [37]. Hence, a mesoscopic treatment with a relatively low computational load is needed. For this, several mesoscopic material models including the Hydrodynamic Drude Model (HDM) [38], [39], [40], its extension Generalized Nonlocal Optical Response (GNOR) model [41], and the Surface Response Model (SRM) [42], [43], have been proposed. The key difference amongst these models lies in the treatment of electron density distribution at the boundary. Both the HDM and the GNOR assume abrupt changes in the charge distribution at an interface (IF), with the HDM considering the electron convection effect, while the GNOR further adding the electron diffusion effect. The SRM looks at the transition region where the electron density distribution varies. The impact of such variation is examined on the level of dipole moments and encoded in the Feibelman parameters [44] and the quantum-corrected boundary conditions [42].

Xin. Zheng is grateful for the China Scholarship Council (Grant No. 202206090027), China. Xuezhi Zheng, and Guy A. E. Vandenbosch are grateful for the C1 project (C14/19/083), the IDN project (IDN/20/014), the small infrastructure grant (KA/20/019) from KU Leuven and for G090017N and G088822N from the Research Foundation of Flanders (FWO). Xuezhi Zheng would like to also thank the support from the IEEE Antennas and Propagation Society Postdoctoral Fellowship, and for the FWO long travel grant, FWO V408823N. The Center for Polariton-driven Light – Matter Interactions (POLIMA) is funded by the Danish National Research Foundation (Project No. DNRF165) (Corresponding author: Xuezhi Zheng).

X. Zheng (e-mail: xin.zheng@esat.kuleuven.be) and G. A. E. Vandenbosch (e-mail: guy.vandenbosch@esat.kuleuven.be) are affiliated with the WaveCore Division, Department of Electrical Engineering (ESAT), KU Leuven, B-3001, Leuven, Belgium.

X. Zheng (e-mail: xuezhi.zheng@esat.kuleuven.be) is affiliated with the Key Laboratory of Radar Imaging and Microwave Photonics of Ministry of Education, Nanjing University of Aeronautics and Astronautics, Nanjing, 211106, Jiangsu, China.

C. Mystilidis (e-mail: chrmys@mci.sdu.dk) and C. Tserkezis (e-mail: ct@mci.sdu.dk) are affiliated with POLIMA – Center for Polariton-driven Light-Matter Interactions, University of Southern Denmark, 5230 Odense, Denmark.

Corresponding to these new developments in semiclassical material models, systematic efforts have been made in the research domain of Computational Electromagnetics (CEM), gradually leading to a new rising direction, i.e., Computational Mesoscopic Electromagnetics (CMEM) [45], [46]. One core task is to integrate classical CEM methods with the mesoscopic material models. Amongst a myriad of topologies, here we solely focus on the cylindrical ones, as they provide a natural abstraction for nanowires (NWs) and can be an ideal platform for studying two-dimensional (2D) quantum and electromagnetic (EM) confinement. For long and thin cylinders, the relatively large circumference-to-area ratio results in magnified nonlocal effects, including spectral shifts, spill-in/out effect [24, p. 7132], [30], electron tunneling through neighboring NWs, and dynamical screening [47]. The HDM has been integrated with the Finite Element Method (FEM) for NW dimers to demonstrate boosted field enhancements [18], and for plasmonic NWs to identify the impact of nonlocal effects on waveguiding properties [48], [49]; with the Finite Difference Time Domain (FDTD) method to trace the nonlocal optical response in cylinders [50], [51]; with the Discontinuous Galerkin Time Domain (DGTD) method to investigate the resonant modes supported in cylindrical dimers [52]; and with the Boundary Element Method (BEM) to study the behavior of plasmonic NWs in different dielectric environments [53]. Further, the GNOR model has been integrated with the DGTD and with the FEM [30] [41] [54]. Regarding the SRM, there have been studies on nanoparticles or layers, but for cylinders a computational tool is still missing. Despite these powerful numerical algorithms, where very complicated geometries can be treated, semi-analytical methods, though limited to canonical shapes (spheres, cylinders and layers), stand out because of their fast computing speed, high accuracy, serving as a perfect (reference) tool for exploring the physics behind. When implementing nonlocal response models, apart from the exploitation of optical properties of simple structures [22], [55], [56], these semi-analytical methods also work on more complicated nanostructures, such as multilayered concentric spheres [57], [58], [59], [60] or cylinders [59], [61], multilayered infinite planes [62], and the nanoparticle on mirror structure [63].

In this paper, we focus on the interaction of light with multiple non-overlapping cylindrical IFs, which may belong to the same or different NWs (see Fig. 1) and develop an efficient algorithm for the problem. The metal constituting the NWs can be treated by HDM (and GNOR) or SRM. Since NWs typically have an optically large extent along the direction perpendicular to their cross-section and are therefore seen as computationally infinitely long [64], we can reduce this problem from a three-dimensional (3D) to a 2D one. The electric and magnetic fields are expanded in terms of vector cylindrical wave functions. In this algorithm, there are two key elements: the scattering matrix (i.e., the S-Matrix) and the use of the translational addition theorem [64]. While the former is dependent on the properties of the materials (and the boundary conditions) and the radius of the NW, the latter links field expansions regarding different expansion centers. Since the **S** matrix plays a pivotal role in the developed algorithm, we refer to the algorithm as an "S-matrix" – based algorithm.

The paper is organized as follows. In Section II, a brief introduction about the HDM, the GNOR and the SRM is given and their corresponding **S** matrix is reviewed. Then, the main equation to be solved by the proposed algorithm is presented, where the discussion emphasizes the use of the translational addition theorem (as discussed before, to reconcile different expansion centers) and the construction of the corresponding matrix. To validate, in Section III, a single gold nanowire is tested, whose scattering cross sections and the near fields from the proposed algorithm are compared with the ones from an in-house developed BEM solver. Here, an excellent agreement is shown. An eccentric dual-material nanowire is further compared to verify the accuracy of the proposed solver for the multiple scattering case. In addition, a physical check on a sodium (Na) nanolens with decreasing radii and gaps is conducted for the LRA, the SRM and the HDM, where the results are in line with the previously reported physical findings for spheres [21]. As a comment, within this work, the SI units are used, and all materials are assumed to be nonmagnetic. The $e^{-i\omega t}$ time dependence is adopted (and for the sake of conciseness, $\omega$ is suppressed throughout the rest of the text).

## II. ALGORITHM

In this section, we review the physical models and the main equation behind the proposed algorithm.

### *A. Physical Models*

To start with, the dynamics of the electric and the magnetic field are controlled by the Maxwell equations in a source-free region,

$$\nabla \times \mathbf{E}(\mathbf{r}) = i\omega\mu_0 \mathbf{H}(\mathbf{r}), \tag{1}$$

$$\nabla \times \mathbf{H}(\mathbf{r}) = -i\omega \mathbf{D}(\mathbf{r}). \tag{2}$$

Here, $\omega$ is the angular frequency and $\mu_0$ stands for the permeability of free space. $\mathbf{E}$ is the electric field; $\mathbf{H}$ is the magnetic field. $\mathbf{D}$ is the electric displacement field, with

$$\mathbf{D}(\mathbf{r}) = \varepsilon_0 \mathbf{E}(\mathbf{r}) + \mathbf{P}(\mathbf{r}). \tag{3}$$

where $\varepsilon_0$ is the vacuum permittivity, and $\mathbf{P}(\mathbf{r})$ is the polarization current, that includes the free-electron part $\mathbf{P}_f(\mathbf{r})$, and bound-electron part $\mathbf{P}_b(\mathbf{r})$. Since the bound electrons are less mobile, $\mathbf{P}_b$ is related with the total electric field $\mathbf{E}$ via $\mathbf{P}_b = \varepsilon_0 \chi_{bd} \mathbf{E}$ where $\chi_{bd}$ is the bound electron susceptibility connected to the relevant permittivity through $\varepsilon_{bd} = 1 + \chi_{bd}$ (see its definition in [65], [66]). Then, for $\mathbf{P}_f$, the HDM and the GNOR are introduced. These two models have been extensively discussed in our previous works [17], [59], [60], [63], [65], [67], [68]. In a nutshell, they are encoded in an additional partial differential equation of the same form,

$$\xi^2(\omega)\nabla\left[\nabla \cdot \mathbf{P}_f(\mathbf{r})\right] + \mathbf{P}_f(\mathbf{r}) = -\varepsilon_0 \frac{\omega_p^2}{\omega(\omega + i\gamma)} \mathbf{E}(\mathbf{r}). \tag{4}$$

For the HDM,

$$\xi^2(\omega) = \frac{\beta^2}{\omega(\omega + i\gamma)}. \tag{5}$$

For the GNOR,

$$\xi^2(\omega) = \frac{\beta^2}{\omega(\omega + i\gamma)} + \frac{\mathfrak{D}}{i\omega}. \tag{6}$$

In the above, $\omega_p$ and $\gamma$ are the plasma frequency and the damping rate of the free electron gas of a metal. For the HDM, $\beta$ is a hydrodynamic parameter related to the Fermi velocity and typically taken to its high-frequency limit $\sqrt{3/5}\upsilon_F$ [32], [69] and the LRA can be achieved by setting $\beta = 0$. While for the GNOR, an additional term $\mathfrak{D}/i\omega$ is included in $\xi^2(\omega)$ to account for the diffusive effects, and $\mathfrak{D}$ is the diffusion constant [70].

Next, consider a source-free region that is homogeneously filled with a material whose EM response can be described by the HDM. The governing wave equations for the region can be obtained by combining the HDM in (4) with Maxwell's equations in (1) and (2), which after Helmholtz decomposition are

$$\nabla \times \nabla \times \mathbf{E}^t(\mathbf{r}, \omega) - k_t{}^2 \mathbf{E}^t(\mathbf{r}, \omega) = 0, \tag{7}$$

$$\nabla^2 \mathbf{E}^l(\mathbf{r}, \omega) + k_l{}^2 \mathbf{E}^l(\mathbf{r}, \omega) = 0. \tag{8}$$

Their derivations can be found in our previous works [65]. Equation (9) describes the dynamics of the transverse waves (denoted by a superscript $t$), while (8) treats the longitudinal waves (marked a superscript $l$). Here, $k_t^2 = (\omega/c)^2\varepsilon(\omega)$ and $k_l^2 = \varepsilon(\omega)/\left(\varepsilon_{bd}(\omega)\xi(\omega)^2\right)$ are the transverse and longitudinal wave numbers, where $\varepsilon(\omega) = \varepsilon_{bd}(\omega) - \omega_p^2/(\omega(\omega + i\gamma))$, and $\varepsilon_{bd}(\omega)$ is the bound electron permittivity.

After discussing a homogeneously filled space, we turn our focus to an IF where interesting physics happens. At an IF, it is well known that apart from the two traditional boundary conditions (the continuity of the tangential components of the electric and magnetic fields),

$$\mathbf{n} \times (\mathbf{E}_1 - \mathbf{E}_2) = \mathbf{0}, \tag{10}$$

$$\mathbf{n} \times (\mathbf{H}_1 - \mathbf{H}_2) = \mathbf{0}, \tag{11}$$

additional boundary conditions (ABCs) are needed, due to the involvement of the longitudinal waves in (8) [59], [71], [72]. On the one hand, for an IF separating a medium described by LRA (i.e., a local medium) and a medium described by HDM (i.e., a nonlocal medium), in accordance with the assumption that the free electron charge density drops abruptly to zero at the boundary in HDM, as an ABC the normal component of the free-electron polarization current is assumed to stop at the IF and this can be expressed in terms of the electric fields as a third BC [73],

$$\mathbf{n} \cdot \varepsilon_{1,bd} \mathbf{E}_1 = \mathbf{n} \cdot \varepsilon_2 \mathbf{E}_2. \tag{12}$$

In (10)-(12), $\mathbf{n}$ is the unit normal of the IF, pointing from the inner region to the outer region; the subscripts 1 and 2 indicate the inner and outer regions separated by the IF; region 1 is filled with a nonlocal medium, $\varepsilon_{1,bd}$ means the bound electron permittivity of the medium in region 1, and region 2 is filled with a local medium with a permittivity $\varepsilon_2$.

On the other hand, for an IF that separates a nonlocal medium from another nonlocal medium, two ABCs [73], [74] are needed,

$$\mathbf{n} \cdot \varepsilon_{1,bd} \mathbf{E}_1 = \mathbf{n} \cdot \varepsilon_{2,bd} \mathbf{E}_2. \tag{13}$$

$$\frac{\beta_1^2}{\omega_{1,p}^2} \varepsilon_{1,bd} \nabla \cdot \mathbf{E}_1 = \frac{\beta_2^2}{\omega_{2,p}^2} \varepsilon_{2,bd} \nabla \cdot \mathbf{E}_2. \tag{14}$$

(14) guarantees the continuity of the normal component of the free electron polarization current, while the divergence part reflects the charge density.

Unlike the HDM and the GNOR focusing on the bulk of a medium, the SRM looks at the field behaviors at an IF and summarizes all the physical details in $d$-parameters, which are also known as the surface response functions that depend on the two materials that make up the IF. In this case, the materials are treated by LRA and only transverse waves exist, referring to Eq. (7). Lying at the core of the SRM are two quantum-corrected boundary conditions (QC-BCs) [55],

$$\mathbf{E}_2^{\parallel} - \mathbf{E}_1^{\parallel} = -d_{\perp} \cdot \nabla_{\parallel}\left(\mathbf{E}_2^{\perp} - \mathbf{E}_1^{\perp}\right). \tag{15}$$

$$\mathbf{H}_2^{\parallel} - \mathbf{H}_1^{\parallel} = -i\omega d_{\parallel} \cdot \mathbf{n} \times \left(\mathbf{D}_2^{\parallel} - \mathbf{D}_1^{\parallel}\right). \tag{16}$$

$d_{\parallel}$ and $d_{\perp}$ are the so-called *Feibelman parameters*, which are dependent on frequency, and can be extracted from, e.g., TD-DFT. The details on these two parameters are explained in [5]. This incorporation of the $d$ parameters in the conventional BCs, i.e., (15) and (16), can be understood by considering their physical impact through (normally-oriented) surface polarization and (tangentially-oriented) current terms [55], which are nonzero only at the IF. By introducing the complex Feibelman parameters, the SRM lumps the quantum mechanical effects, e.g., electron spill-out and surface enhanced Landau damping, that are prominent close to the boundary.

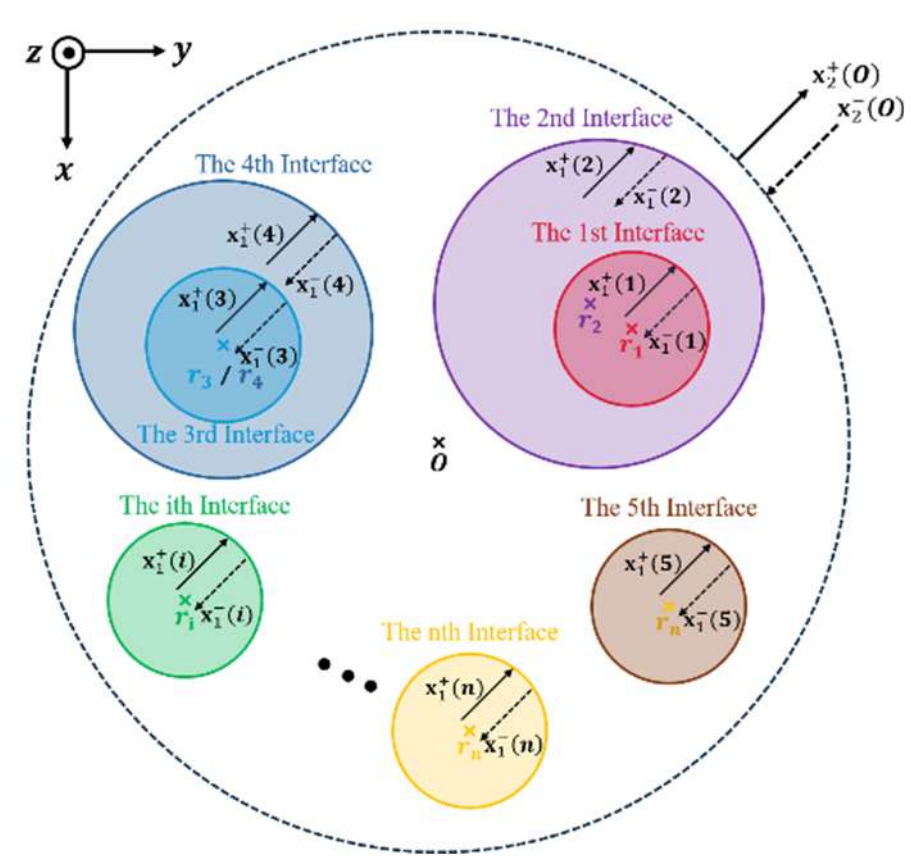

Fig. 1. A cross-sectional view of $N$ interacting nanowires. There are $N$ IFs in total. The subscripts 1 and 2 mark the inner and outer region of each IF, respectively. Corresponding to the superscripts "+" and "-", the arrows indicate whether the waves are converging or diverging with respect to their own origins. The dashed circle outside all nanowires is an imaginary boundary collecting all the scattered and incident waves. In the figure, IF1 and IF2 constitute an eccentric nanowire, while IF3 and IF4 form a concentric nanowire.

*B. Single Cylindrical Interface: S-Matrix*

We first consider a standalone infinitely long nanowire in vacuum. For this, we look at the general solutions to the wave equations in (7) and (8), which are known as the cylindrical vector wave functions (CVWFs) $\mathbf{M}_m(\mathbf{r})$, $\mathbf{N}_m(\mathbf{r})$ and $\mathbf{L}_m(\mathbf{r})$ [64],

$$\mathbf{M}_m(\mathbf{r}) = \nabla\times\left[\psi_m^t(\mathbf{r})\mathbf{z}\right] = \frac{im}{\rho}\psi_m^t(\mathbf{r})\hat{\boldsymbol{\rho}} - \frac{\partial\psi_m^t(\mathbf{r})}{\partial\rho}\hat{\boldsymbol{\varphi}}, \tag{17}$$

$$\begin{aligned}\mathbf{N}_m(\mathbf{r}) &= \frac{1}{k}\nabla\times\nabla\times\left[\psi_m^t(\mathbf{r})\mathbf{z}\right] \\ &= \frac{ik_z}{k}\frac{\partial\psi_m^t(\mathbf{r})}{\partial\rho}\hat{\boldsymbol{\rho}} + \frac{ik_z}{k}\frac{im}{\rho}\psi_m^t(\mathbf{r})\hat{\boldsymbol{\varphi}} + \frac{k_\rho^2}{k}\psi_m^t(\mathbf{r})\hat{\mathbf{z}},\end{aligned} \tag{18}$$

$$\mathbf{L}_m(\mathbf{r}) = \nabla\psi_m^l(\mathbf{r}) = \frac{\partial\psi_m^l(\mathbf{r})}{\partial\rho}\hat{\boldsymbol{\rho}} + \frac{im}{\rho}\psi_m^l(\mathbf{r})\hat{\boldsymbol{\varphi}} + ik_z\psi_m^l(\mathbf{r})\hat{\mathbf{z}}. \tag{19}$$

In the above, $\psi(\mathbf{r})$ is the scalar cylindrical wave function and the superscripts $t$ and $l$ mark the use of transverse and longitudinal wavenumbers $k_\rho$ and $\kappa_\rho$,

$$\psi_m^t(\mathbf{r}) = Z_m(k_\rho\rho)e^{im\varphi}e^{ik_z z}, \tag{20}$$

$$\psi_m^l(\mathbf{r}) = Z_m(\kappa_\rho\rho)e^{im\varphi}e^{ik_z z}. \tag{21}$$

In detail, the transverse cylindrical wave is indexed by three parameters $k_\rho$, $k_z$ and $m$, where $k_\rho$ and $k_z$ are linked by $k_\rho = \sqrt{k_t{}^2 - k_z^2}$, and $m$ is the order of the cylindrical wave. In principle, the value $m$ can be an integer from $-\infty$ to $+\infty$, while the corresponding parameters for longitudinal waves are $\kappa_\rho$, $k_z$ and $m$, where $\kappa_\rho$ and $k_z$ are linked by $\kappa_\rho = \sqrt{k_l{}^2 - k_z^2}$. For both cases, $k_z$ is assumed to be fixed, e.g., by an external incident field. The $Z$ function can be Bessel functions $J_m$ (known as the "standing" wave type) or Hankel functions of the first kind $H_m^{(1)}$ (known as the "radiating" wave type).

For the cylindrical IF centered at $\mathbf{r}_p$, the electric field in the inside or outside the IF can be expanded in terms of CVWFs. Here, we consider the most complicated case where the region is filled with a nonlocal medium,

$$\begin{aligned}\mathbf{E}(\mathbf{r}) &= \sum_m \mathbf{M}_m^{(1/3)}\left(k_\rho, k_z, \mathbf{r}-\mathbf{r}_p\right)\cdot a_m^{(1/3)} \\ &+ \sum_m \mathbf{N}_m^{(1/3)}\left(k_\rho, k_z, \mathbf{r}-\mathbf{r}_p\right)\cdot b_m^{(1/3)} \\ &+ \sum_m \mathbf{L}_m^{(1/3)}\left(\kappa_\rho, k_z, \mathbf{r}-\mathbf{r}_p\right)\cdot c_m^{(1/3)}.\end{aligned} \tag{22}$$

$a_m$, $b_m$ and $c_m$, are the expansion coefficients of the CVWFs. The longitudinal wave function $\mathbf{L}_m$ and the corresponding expansion coefficient $c_m$ should be ignored when considering a local medium. For the superscripts, taking (1) indicates that the wave function is of the standing (converging) wave type, related to Bessel function; while taking (3) means that the wave function is of the radiating (diverging) wave type, related to Hankel function.

In the inner or outer region of each IF, the field is the sum of two components: the diverging waves and the converging waves. And the expansion coefficients $a_m^{(1/3)}$, $b_m^{(1/3)}$ and $c_m^{(1/3)}$ of these two kinds of waves are linked by the $\mathbf{S}$ matrix of the IF, which is obtained by matching the fields inside and outside the IF. The construction for the $\mathbf{S}$ matrix has been introduced in our previous work [59] and here we only review the key points. For the purpose of a clear and intuitive demonstration, first we focus on a single $m$. The fields on both sides of the IF are connected by the BCs (i.e., (10) - (14) for the HDM and (15) - (16) for the SRM) and then projected onto the unit vectors of the cylindrical coordinate system, i.e., $\hat{\boldsymbol{\rho}}$, $\hat{\boldsymbol{\varphi}}$, and $\hat{\mathbf{z}}$ (the unit vectors along the radial, azimuthal and z directions). As a result, we obtain the $\mathbf{S}$ matrices,

$$\mathbf{S}_m = \left[\mathbf{M}_1^{(3)} \quad -\mathbf{M}_2^{(1)}\right]^{-1}\cdot\left[-\mathbf{M}_1^{(1)} \quad \mathbf{M}_2^{(3)}\right]. \tag{23}$$

In (23), $\mathbf{M}$ are matrices whose detailed mathematical forms are shown below. Their superscripts can take the value of (1) or (3), deciding whether the Bessel or the Hankel function is used in the matrices. Their subscripts can be 1 or 2 telling if the inner or the outer region of the IF is considered. This will immediately affect the wavenumber in the argument of the Bessel or the Hankel functions.

**First**, for an IF separating two local media, the conventional boundary conditions in (10) and (11) are applied to guarantee the continuity of the tangential components of the electric and magnetic field. Projecting these fields on the $\hat{\mathbf{z}}$ and $\hat{\boldsymbol{\varphi}}$ axis, the **M** matrix reads

$$\mathbf{M}=\begin{pmatrix} 0 & \frac{k_\rho^2}{k_t}Z_m\left(k_\rho\rho\right) \\ \frac{1}{iZ}\cdot\frac{k_\rho^2}{k_t}Z_m\left(k_\rho\rho\right) & 0 \\ -\frac{\partial Z_m\left(k_\rho\rho\right)}{\partial\rho} & \frac{ik_z}{k_t}\frac{im}{\rho}Z_m\left(k_\rho\rho\right) \\ \frac{1}{iZ}\cdot\frac{ik_z}{k_t}\frac{im}{\rho}Z_m\left(k_\rho\rho\right) & \frac{1}{iZ}\cdot-\frac{\partial Z_m\left(k_\rho\rho\right)}{\partial\rho} \end{pmatrix} \tag{24}$$

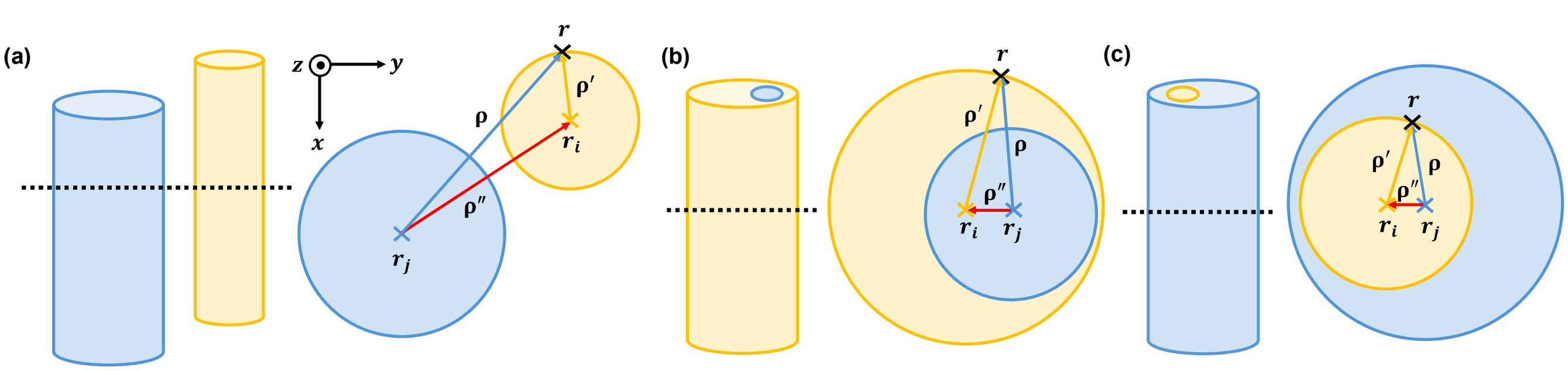

Fig. 2. An illustration of three different cases when two IFs share a common region. The "source" cylinder is indicated in blue, while the "observation" cylinder is indicated in yellow. The black dotted line indicates a transverse cross-section of the structure, and the cross-sectional view is shown on the right side. The black cross indicates an observation point $r$. (a) case 1: the two cylinders are outside of each other; (b) case 2: the "source" cylinder is inside the "observation" cylinder; and (c) case 3: the "observation" cylinder is inside the "source" cylinder.

Here, $Z$ represents the wave impedance. If we consider the $\mathbf{M}^{(1)}$, $Z_m$ should be a Bessel function; while for $\mathbf{M}^{(3)}$, a Hankel function is used. Also, if the inner (outer) region is considered, $k_\rho$ and $k_t$ of the inner (outer) region should be used.

**Second**, when the IF separates a nonlocal medium (inner) from a local medium (outer), since the amplitude of longitudinal waves is additionally needed, the ABC in (12) should be considered. For the inner region, $\mathbf{M}_1$ matrix reads,

$$\mathbf{M}_1=\begin{pmatrix} 0 & +\frac{k_\rho^2}{k_{1,t}}Z_m\left(k_\rho\rho\right) & +ik_zZ_m\left(\kappa_\rho\rho\right) \\ \frac{1}{iZ}\cdot\frac{k_\rho^2}{k_{1,t}}Z_m\left(k_\rho\rho\right) & 0 & 0 \\ -\frac{\partial Z_m\left(k_\rho\rho\right)}{\partial\rho} & \frac{ik_z}{k_{1,t}}\frac{im}{\rho}Z_m\left(k_\rho\rho\right) & +\frac{im}{\rho}Z_m\left(\kappa_\rho\rho\right) \\ \frac{1}{iZ}\cdot\frac{ik_z}{k_{1,t}}\frac{im}{\rho}Z_m\left(k_\rho\rho\right) & \frac{1}{iZ}\cdot-\frac{\partial Z_m\left(k_\rho\rho\right)}{\partial\rho} & 0 \\ \varepsilon_{1,bd}\frac{im}{\rho}Z_m\left(k_\rho\rho\right) & \varepsilon_{1,bd}\frac{ik_z}{k_{1,t}}\frac{\partial Z_m\left(k_\rho\rho\right)}{\partial\rho} & \varepsilon_{1,bd}\frac{\partial Z_m\left(\kappa_\rho\rho\right)}{\partial\rho} \end{pmatrix}. \tag{25}$$

For the outer region, the third column in the $\mathbf{M}_2$ matrix should be ignored,

$$\mathbf{M}_2=\begin{pmatrix} 0 & +\frac{k_\rho^2}{k_{2,t}}Z_m\left(k_\rho\rho\right) \\ \frac{1}{iZ}\cdot\frac{k_\rho^2}{k_{2,t}}Z_m\left(k_\rho\rho\right) & 0 \\ -\frac{\partial Z_m\left(k_\rho\rho\right)}{\partial\rho} & \frac{ik_z}{k_{2,t}}\frac{im}{\rho}Z_m\left(k_\rho\rho\right) \\ \frac{1}{iZ}\cdot\frac{ik_z}{k_{2,t}}\frac{im}{\rho}Z_m\left(k_\rho\rho\right) & \frac{1}{iZ}\cdot-\frac{\partial Z_m\left(k_\rho\rho\right)}{\partial\rho} \\ \varepsilon_2\frac{im}{\rho}Z_m\left(k_\rho\rho\right) & \varepsilon_2\frac{ik_z}{k_{2,t}}\frac{\partial Z_m\left(k_\rho\rho\right)}{\partial\rho} \end{pmatrix}. \tag{26}$$

Once again, if the superscript of $\mathbf{M}_1$ and $\mathbf{M}_2$ is (1) (or (3)), $Z_m$ should take the Bessel (Hankel) function.

**Third,** the IF separates a nonlocal medium from a nonlocal medium. The fields on both sides of the IF are matched by the BCs in (10), (11), (13) and (14). By further projecting onto the unit vectors, we reach (27),

$$\mathbf{M}=\begin{pmatrix} 0 & +\frac{k_\rho^2}{k_t}Z_m\left(k_\rho\rho\right) & +ik_zZ_m\left(\kappa_\rho\rho\right) \\ \frac{1}{iZ}\cdot\frac{k_\rho^2}{k_t}Z_m\left(k_\rho\rho\right) & 0 & 0 \\ -\frac{\partial Z_m\left(k_\rho\rho\right)}{\partial\rho} & +\frac{ik_z}{k_t}\frac{im}{\rho}Z_m\left(k_\rho\rho\right) & +\frac{im}{\rho}Z_m\left(\kappa_\rho\rho\right) \\ \frac{1}{iZ}\cdot\frac{ik_z}{k_t}\frac{im}{\rho}Z_m\left(k_\rho\rho\right) & \frac{1}{iZ}\cdot-\frac{\partial Z_m\left(k_\rho\rho\right)}{\partial\rho} & 0 \\ \varepsilon_{bd}\frac{im}{\rho}Z_m\left(k_\rho\rho\right) & \varepsilon_{bd}\frac{ik_z}{k_t}\frac{\partial Z_m\left(k_\rho\rho\right)}{\partial\rho} & \varepsilon_{bd}\frac{\partial Z_m\left(\kappa_\rho\rho\right)}{\partial\rho} \\ 0 & 0 & -\frac{\beta^2}{\omega_p^2}\varepsilon_{bd}k_l^2Z_m\left(\kappa_\rho\rho\right) \end{pmatrix}. \tag{27}$$

**Lastly**, if the SRM is considered, the fields on both sides of the IF should satisfy (15) and (16) instead, which leads to,

$$\mathbf{M}=\begin{pmatrix} 0 & +\frac{k_\rho^2}{k_t}Z_m\left(k_\rho\rho\right) & +ik_zZ_m\left(\kappa_\rho\rho\right) \\ \frac{1}{iZ}\cdot\frac{k_\rho^2}{k_t}Z_m\left(k_\rho\rho\right) & 0 & 0 \\ -\frac{\partial Z_m\left(k_\rho\rho\right)}{\partial\rho} & +\frac{ik_z}{k_t}\frac{im}{\rho}Z_m\left(k_\rho\rho\right) & +\frac{im}{\rho}Z_m\left(\kappa_\rho\rho\right) \\ \frac{1}{iZ}\cdot\frac{ik_z}{k_t}\frac{im}{\rho}Z_m\left(k_\rho\rho\right) & \frac{1}{iZ}\cdot-\frac{\partial Z_m\left(k_\rho\rho\right)}{\partial\rho} & 0 \\ \varepsilon_{bd}\frac{im}{\rho}Z_m\left(k_\rho\rho\right) & \varepsilon_{bd}\frac{ik_z}{k_t}\frac{\partial Z_m\left(k_\rho\rho\right)}{\partial\rho} & \varepsilon_{bd}\frac{\partial Z_m\left(\kappa_\rho\rho\right)}{\partial\rho} \\ 0 & 0 & -\frac{\beta^2}{\omega_p^2}\varepsilon_{bd}k_l^2Z_m\left(\kappa_\rho\rho\right) \end{pmatrix}. \tag{28}$$

In (26) - (28), if the superscript of the **M** matrices is (1) (or (3)), $Z_m$ should take the Bessel (Hankel) function. Theoretically, the expansion involves all integer orders $m$, and $m \in \mathbb{Z}$. In practice, it is truncated to $-M \le m \le M$, where $M$ is the cutoff parameter. Therefore, there are in total $(2M+1)$ $a_m$, $b_m$ and $c_m$ which can be put into column vectors,

$$\begin{aligned} \mathbf{a}&=\left(a_{-M}\ \ \dots\ \ a_0\ \ \dots\ \ a_{+M}\right)^T, \\ \mathbf{b}&=\left(b_{-M}\ \ \dots\ \ b_0\ \ \dots\ \ b_{+M}\right)^T, \\ \mathbf{c}&=\left(c_{-M}\ \ \dots\ \ c_0\ \ \dots\ \ c_{+M}\right)^T. \end{aligned} \tag{29}$$

For the sake of conciseness, we summarize **a**, **b** and **c** in **x**,

$$\mathbf{x}=\left(\mathbf{a}\quad\mathbf{b}\quad\mathbf{c}\right)^T. \tag{30}$$

Thus, the **S** matrix, introduced by the **M** matrices as in (23) and linking the expansion coefficients of CVWFs, is expressed as,

$$\begin{pmatrix}\mathbf{x}_1^-\\ \mathbf{x}_2^+\end{pmatrix}=\mathbf{S}\cdot\begin{pmatrix}\mathbf{x}_1^+\\ \mathbf{x}_2^-\end{pmatrix}. \tag{31}$$

The superscripts "+" and "-" correspond to the converging and diverging waves.

### *C. Translational Addition Theorem*

When considering the scattering of multiple cylinders we need to reconcile the wave expansions with respect to different expansion centers. To be specific, we consider the interaction of two cylinders. One of the two cylinders is called "source" cylinder which is centered at $\mathbf{r}_j$, while the other is called "observation" cylinder centered at $\mathbf{r}_i$. In the first case, the outer region of cylinder $i$ is the same as the outer region of cylinder $j$, shown in Fig. 2(a). The electric field at an observation point $\mathbf{r}$ can be expanded in terms of cylindrical waves centered at the "source" cylinder,

$$\mathbf{E}(\mathbf{r}) = \sum_m \begin{matrix} \mathbf{M}_m^{(3)}\left(k_\rho, k_z, \mathbf{r}-\mathbf{r}_j\right)\cdot a_m(j) + \\ \mathbf{N}_m^{(3)}\left(k_\rho, k_z, \mathbf{r}-\mathbf{r}_j\right)\cdot b_m(j) + \\ \mathbf{L}_m^{(3)}\left(\kappa_\rho, k_z, \mathbf{r}-\mathbf{r}_j\right)\cdot c_m(j). \end{matrix} \tag{32}$$

In (32), the radiating wave type of CVWFs is used in the expansion. Further, the same electric field at the observation point $\mathbf{r}$ can be expanded in terms of cylindrical waves centered at the "observation" cylinder,

$$\mathbf{E}(\mathbf{r}) = \sum_n \begin{matrix} \mathbf{M}_n^{(1)}\left(k_\rho, k_z, \mathbf{r}-\mathbf{r}_i\right)\cdot a_n(i) + \\ \mathbf{N}_n^{(1)}\left(k_\rho, k_z, \mathbf{r}-\mathbf{r}_i\right)\cdot b_n(i) + \\ \mathbf{L}_n^{(1)}\left(\kappa_\rho, k_z, \mathbf{r}-\mathbf{r}_i\right)\cdot c_n(i). \end{matrix} \tag{33}$$

In (33), the standing wave type of CVWFs is used in the expansion. Since we always consider the case where $z$ is unchanged, the displacement can be written as $\boldsymbol{\rho} = \boldsymbol{\rho}' + \boldsymbol{\rho}''$. For $\mathbf{M}_m^{(3)}(k_\rho, k_z, \mathbf{r})$, $\mathbf{N}_m^{(3)}(k_\rho, k_z, \mathbf{r})$ and $\mathbf{L}_m^{(3)}(\kappa_\rho, k_z, \mathbf{r})$, in whose expressions the Hankel function of the first kind is used, when $\boldsymbol{\rho}''$ is larger than $\boldsymbol{\rho}'$, according to translational addition theorem, we have,

$$\begin{aligned} \mathbf{M}_m^{(3)}\left(k_\rho, k_z, \mathbf{r}\right) &= \sum_{n=-\infty}^{+\infty} H_{m-n}^{(1)}\left(k_\rho \rho''\right) e^{i(m-n)\varphi''} \cdot \mathbf{M}_n^{(1)}\left(k_\rho, k_z, \mathbf{r}'\right), \\ \mathbf{N}_m^{(3)}\left(k_\rho, k_z, \mathbf{r}\right) &= \sum_{n=-\infty}^{+\infty} H_{m-n}^{(1)}\left(k_\rho \rho''\right) e^{i(m-n)\varphi''} \cdot \mathbf{N}_n^{(1)}\left(k_\rho, k_z, \mathbf{r}'\right), \\ \mathbf{L}_m^{(3)}\left(\kappa_\rho, k_z, \mathbf{r}\right) &= \sum_{n=-\infty}^{+\infty} H_{m-n}^{(1)}\left(\kappa_\rho \rho''\right) e^{i(m-n)\varphi''} \cdot \mathbf{L}_n^{(1)}\left(\kappa_\rho, k_z, \mathbf{r}'\right). \end{aligned} \tag{34}$$

Then, we note that,

$$\begin{aligned} a_n(i) &= \sum_m H_{m-n}^{(1)}\left(k_\rho \rho''\right) e^{i(m-n)\varphi''} \cdot a_m(j), \\ b_n(i) &= \sum_m H_{m-n}^{(1)}\left(k_\rho \rho''\right) e^{i(m-n)\varphi''} \cdot b_m(j), \\ c_n(i) &= \sum_m H_{m-n}^{(1)}\left(\kappa_\rho \rho''\right) e^{i(m-n)\varphi''} \cdot c_m(j). \end{aligned} \tag{35}$$

In the second case, the outer region of the "source" cylinder is the same as the inner region of the "observation" cylinder (see Fig. 2(b)), the electric field at the observation point $\mathbf{r}$ can be expanded regarding the "source" cylinder,

$$\mathbf{E}(\mathbf{r}) = \sum_m \begin{matrix} \mathbf{M}_m^{(3)}\left(k_\rho, k_z, \mathbf{r}-\mathbf{r}_j\right)\cdot a_m(j) + \\ \mathbf{N}_m^{(3)}\left(k_\rho, k_z, \mathbf{r}-\mathbf{r}_j\right)\cdot b_m(j) + \\ \mathbf{L}_m^{(3)}\left(\kappa_\rho, k_z, \mathbf{r}-\mathbf{r}_j\right)\cdot c_m(j). \end{matrix} \tag{36}$$

Or the "observation" cylinder,

$$\mathbf{E}(\mathbf{r}) = \sum_n \begin{matrix} \mathbf{M}_n^{(1)}\left(k_\rho, k_z, \mathbf{r}-\mathbf{r}_i\right)\cdot a_n(i) + \\ \mathbf{N}_n^{(1)}\left(k_\rho, k_z, \mathbf{r}-\mathbf{r}_i\right)\cdot b_n(i) + \\ \mathbf{L}_n^{(1)}\left(\kappa_\rho, k_z, \mathbf{r}-\mathbf{r}_i\right)\cdot c_n(i). \end{matrix} \tag{37}$$

When $\boldsymbol{\rho}''$ is always smaller than $\boldsymbol{\rho}'$, we have,

$$\begin{aligned} \mathbf{M}_m^{(3)}\left(k_\rho, k_z, \mathbf{r}\right) &= \sum_{n=-\infty}^{+\infty} J_{m-n}\left(k_\rho \rho''\right) e^{i(m-n)\varphi''} \cdot \mathbf{M}_n^{(3)}\left(k_\rho, k_z, \mathbf{r}'\right), \\ \mathbf{N}_m^{(3)}\left(k_\rho, k_z, \mathbf{r}\right) &= \sum_{n=-\infty}^{+\infty} J_{m-n}\left(k_\rho \rho''\right) e^{i(m-n)\varphi''} \cdot \mathbf{N}_n^{(3)}\left(k_\rho, k_z, \mathbf{r}'\right), \\ \mathbf{L}_m^{(3)}\left(\kappa_\rho, k_z, \mathbf{r}\right) &= \sum_{n=-\infty}^{+\infty} J_{m-n}\left(\kappa_\rho \rho''\right) e^{i(m-n)\varphi''} \cdot \mathbf{L}_n^{(3)}\left(\kappa_\rho, k_z, \mathbf{r}'\right). \end{aligned} \tag{38}$$

Then, we can obtain,

$$\begin{aligned} a_n(i) &= \sum_n J_{m-n}\left(k_\rho \rho''\right) e^{i(m-n)\varphi''} \cdot a_m(j), \\ b_n(i) &= \sum_n J_{m-n}\left(k_\rho \rho''\right) e^{i(m-n)\varphi''} \cdot b_m(j), \\ c_n(i) &= \sum_n J_{m-n}\left(\kappa_\rho \rho''\right) e^{i(m-n)\varphi''} \cdot c_m(j). \end{aligned} \tag{39}$$

In the third case, the inner region of the "source" cylinder is the same as the outer region of the "observation" cylinder (see Fig. 2(c)), the electric field at the observation point $\mathbf{r}$ can be expanded about the "source" cylinder,

$$\mathbf{E}(\mathbf{r}) = \sum_m \begin{matrix} \mathbf{M}_m^{(1)}\left(k_\rho, k_z, \mathbf{r}-\mathbf{r}_j\right)\cdot a_m(j) + \\ \mathbf{N}_m^{(1)}\left(k_\rho, k_z, \mathbf{r}-\mathbf{r}_j\right)\cdot b_m(j) + \\ \mathbf{L}_m^{(1)}\left(\kappa_\rho, k_z, \mathbf{r}-\mathbf{r}_j\right)\cdot c_m(j). \end{matrix} \tag{40}$$

or the "observation" cylinder,

$$\mathbf{E}(\mathbf{r}) = \sum_n \begin{matrix} \mathbf{M}_n^{(1)}\left(k_\rho, k_z, \mathbf{r}-\mathbf{r}_i\right)\cdot a_n(i) + \\ \mathbf{N}_n^{(1)}\left(k_\rho, k_z, \mathbf{r}-\mathbf{r}_i\right)\cdot b_n(i) + \\ \mathbf{L}_n^{(1)}\left(\kappa_\rho, k_z, \mathbf{r}-\mathbf{r}_i\right)\cdot c_n(i). \end{matrix} \tag{41}$$

The standing wave type of CVWFs is used in both expansions, for $\mathbf{M}_m^{(1)}(k_\rho, k_z, \mathbf{r})$, $\mathbf{N}_m^{(1)}(k_\rho, k_z, \mathbf{r})$ and $\mathbf{L}_m^{(1)}(\kappa_\rho, k_z, \mathbf{r})$, in whose expressions the Bessel function is used, and we have

$$\begin{aligned} \mathbf{M}_m^{(1)}\left(k_\rho, k_z, \mathbf{r}\right) &= \sum_{n=-\infty}^{+\infty} J_{m-n}\left(k_\rho \rho''\right) e^{i(m-n)\varphi''} \cdot \mathbf{M}_n^{(1)}\left(k_\rho, k_z, \mathbf{r}'\right), \\ \mathbf{N}_m^{(1)}\left(k_\rho, k_z, \mathbf{r}\right) &= \sum_{n=-\infty}^{+\infty} J_{m-n}\left(k_\rho \rho''\right) e^{i(m-n)\varphi''} \cdot \mathbf{N}_n^{(1)}\left(k_\rho, k_z, \mathbf{r}'\right), \\ \mathbf{L}_m^{(1)}\left(\kappa_\rho, k_z, \mathbf{r}\right) &= \sum_{n=-\infty}^{+\infty} J_{m-n}\left(\kappa_\rho \rho''\right) e^{i(m-n)\varphi''} \cdot \mathbf{L}_n^{(1)}\left(\kappa_\rho, k_z, \mathbf{r}'\right). \end{aligned} \tag{42}$$

Then we get,

$$a_n(i)=\sum_m H^{(1)}_{m-n}\left(k_\rho\rho''\right)e^{i(m-n)\varphi''}\cdot a_m(j),$$
$$b_n(i)=\sum_m H^{(1)}_{m-n}\left(k_\rho\rho''\right)e^{i(m-n)\varphi''}\cdot b_m(j),\qquad(43)$$
$$c_n(i)=\sum_m H^{(1)}_{m-n}\left(\kappa_\rho\rho''\right)e^{i(m-n)\varphi''}\cdot c_m(j).$$

Under no circumstances does the inner region of the "source" cylinder overlap with the inner region of the "observation" cylinder. Therefore, to sum up, a general relation that links the expansion coefficients of the CVWFs "leaving" the $j^{\text{th}}$ interface with the ones "incident" on the $i^{\text{th}}$ interface, is given,

$$\begin{pmatrix}\mathbf{x}_1^{s,+}(i)\\ \mathbf{x}_2^{s,-}(i)\end{pmatrix}=\begin{pmatrix}\mathbf{0} & \mathbf{T}_{12}\\ \mathbf{T}_{21} & \mathbf{T}_{22}\end{pmatrix}\cdot\begin{pmatrix}\mathbf{x}_1^{-}(j)\\ \mathbf{x}_2^{+}(j)\end{pmatrix}.\qquad(44)$$

The subscripts 1and 2 refer to the inner region and the outer region of an IF, respectively. $\mathbf{T}_{22}$, $\mathbf{T}_{12}$, and $\mathbf{T}_{21}$ are block diagonal matrices. The term on the diagonal for $\mathbf{T}_{22}$, $\mathbf{T}_{12}$, and $\mathbf{T}_{21}$ can be deduced from (35), (39) and (43), respectively.

### *D. Main Equation*

Finally, we list the main equation behind our algorithm. Lying at the core of the algorithm is the $\mathbf{S}$ matrix, linking the expansion coefficients of the incoming cylindrical waves with the ones of the outgoing cylindrical waves. The $\mathbf{S}$ matrix can be obtained by (23) considering different boundary conditions. First, we focus on the $u^{\text{th}}$ IF,

$$\begin{pmatrix}\mathbf{x}_1^{-}(u)\\ \mathbf{x}_2^{+}(u)\end{pmatrix}=\mathbf{S}(u)\cdot\begin{pmatrix}\mathbf{x}_1^{+}(u)\\ \mathbf{x}_2^{-}(u)\end{pmatrix}.\qquad(45)$$

The right part, the incident field, is composed of two parts, the first part is the direct incident field coming from the excitation, which is the source of the operating system (plane waves) and is known; the second part is all the scattered fields coming from other $N-1$ interfaces, linked by the translation matrix $\mathbf{T}$, applying (44),

$$\begin{pmatrix}\mathbf{x}_1^{+}(u)\\ \mathbf{x}_2^{-}(u)\end{pmatrix}=\begin{pmatrix}\mathbf{x}_1^{i,+}(u)\\ \mathbf{x}_2^{i,-}(u)\end{pmatrix}+\sum_{v=1,v\neq u}^{N}\begin{pmatrix}\mathbf{0} & \mathbf{T}_{12}\\ \mathbf{T}_{21} & \mathbf{T}_{22}\end{pmatrix}\cdot\begin{pmatrix}\mathbf{x}_1^{-}(v)\\ \mathbf{x}_2^{+}(v)\end{pmatrix}.\qquad(46)$$

Then, we substitute (46) into (45), and shuffle the terms on both sides, to get the main equation of the solver,

$$\begin{pmatrix}\mathbf{x}_1^{-}(u)\\ \mathbf{x}_2^{+}(u)\end{pmatrix}-\mathbf{S}(u)\cdot\sum_{v=1,v\neq u}^{N}\begin{pmatrix}\mathbf{0} & \mathbf{T}_{12}\\ \mathbf{T}_{21} & \mathbf{T}_{22}\end{pmatrix}\cdot\begin{pmatrix}\mathbf{x}_1^{-}(v)\\ \mathbf{x}_2^{+}(v)\end{pmatrix}=\mathbf{S}(u)\cdot\begin{pmatrix}\mathbf{x}_1^{i,+}(u)\\ \mathbf{x}_2^{i,-}(u)\end{pmatrix}.\qquad(47)$$

In the same way, the equations for all the other $N-1$ IFs can be listed. By solving this system of equations, the expansion coefficients of the scattered fields are calculated and thus the scattered field can be recovered.

## III. RESULTS

In this section, we demonstrate numerical results from the proposed algorithm. Three examples are introduced. The first is a single infinitely long gold nanowire (shown in the inset of Fig. 3(a)); the second is an eccentric dual-material nanowire (shown in the inset of Fig. 4(a)); and the third is a self-similar cylindrical nanolens (shown in Fig 7). For the first two examples, we compare the results from the proposed algorithm with data generated by an in-house developed BEM solver [17], while the third one is discussed from a physical perspective. The algorithm is coded in MATLAB, and the computation is performed on a computer with AMD Ryzen 7 PRO 5850U 1.90 GHz and 16 GB RAM. For the sake of completeness, when solving the system of Eq. (47), we did preconditioning for the system matrix by using "equilibrate" with the GMRES iterative solver in MATLAB, so that a quick convergence can be reached.

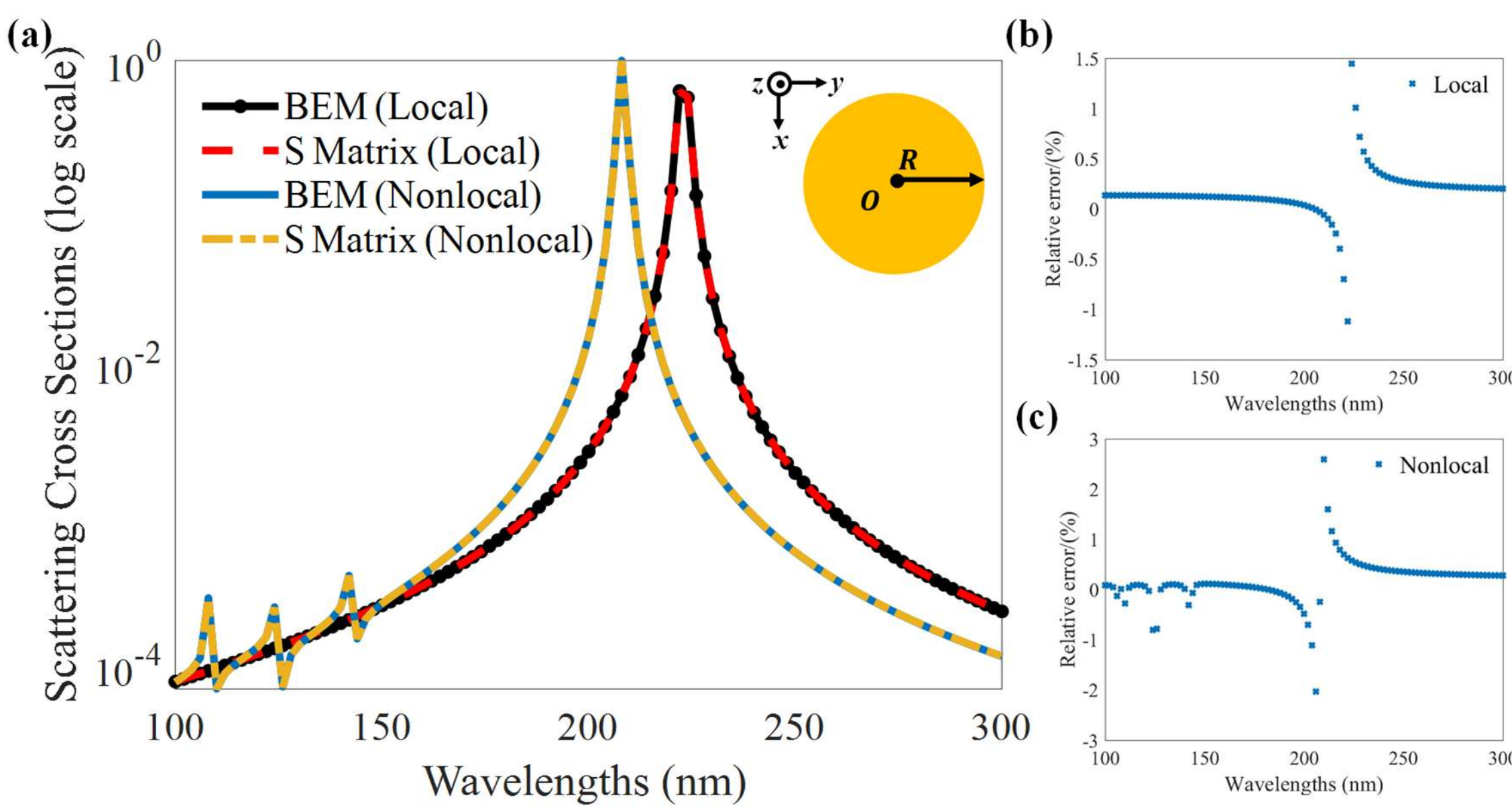

Fig. 3. Simulation results for a gold nanowire. (a) shows the scattering cross sections in a log (common logarithmic) scale. The inset of (a) shows the cross section of the nanowire in the $xy$ plane. For the LRA case, the cross sections are calculated by using the BEM (black dotted line) and proposed algorithm (red dashed line). For the HDM case, the cross sections are done by the BEM (blue line) and the proposed algorithm (yellow dashed line). In (b) and (c), the relative errors as defined in (48) are shown.

### *A. A Single Nanowire*

First, we consider a single nanowire as an example (see the cross section of the wire in Fig. 3(a)). The structure is placed in vacuum and has one cylindrical IF. The IF separates gold (Au, the inner region) from vacuum (the outer region). The center of the outer IF is set as the origin $O$ of the coordinate system and its radius is $R_1 = 1$ nm. Here, we consider two cases: 1) Au is treated as a local medium (LRA); 2) Au is treated as a nonlocal medium (HDM). For the HDM, the Drude damping rate of Au is $8.05 \times 10^{13}$ rad/s, taken from tabulated data [75], and the Fermi velocity of Au is $1.40 \times 10^{6}$ m/s [76]. Here, the effective plasma frequency entering the Drude term is defined as $\sqrt{f_0}\omega_p$, where $f_0 = 0.760$ is the oscillator strength of the free-electron Drude contribution [75]. This gives an effective plasma wavelength of approximately 157.5 nm. The structure is excited by a transverse electric (TE) plane wave travelling along the positive $x$ direction and polarized along the $y$ axis. The amplitude of the incident electric field is 1 V/m. We consider a wavelength range from 100 nm to 300 nm with 101 sampling points taken in between. As the result of the simulation, the scattering cross sections $\sigma_{sca}$ are evaluated by integrating the scattered Poynting flux on a circle $\sigma_{sca} = Z_0 \oint Re(\mathbf{E}_{sca} \times \mathbf{H}^*_{sca}) \cdot \mathbf{n}\, dl.$ $Z_0$ is the free-space impedance and $\mathbf{n}$ is the outward unit normal vector. We consider a circle with a radius of 1 m, which is discretized into 88 line segments on the $xy$ plane. The electric field, the magnetic field and thus the Poynting vector are evaluated at the centers of the line segments. The scattered power (and the cross sections as the ratio of the scattered power to the power density of the incident field) are obtained by numerical integrations.

We consider two cases: 1) Au is treated as a local medium; and 2) Au is treated as a nonlocal medium. The obtained scattering cross sections are shown in Fig 3(a). The results obtained by the proposed algorithm agree well with the ones predicted by the BEM, and the difference is evaluated by the relative error,

$$err = \frac{\sigma^{S}_{sca} - \sigma^{BEM}_{sca}}{\sigma^{S}_{sca}} \quad .(48)$$

$\sigma^{S}_{sca}$ represents the scattering cross sections obtained by using the proposed **S** Matrix Method, while $\sigma^{BEM}_{sca}$ represents the scattering cross sections obtained by using BEM. The maximum relative error is within 3% (shown in Fig. 3(b-c)) and is seen at the resonant wavelength. When using the HDM (compared to the LRA case), a blue-shift of the main resonance and three extra resonances at the high-frequency band can be clearly observed in Fig. 3(a). The blue shift is due to the “spill-in” nature of the volume charge, that is, the HDM effectively reduces the electrical size of the cylindrical IF. The extra peaks are the results of the longitudinal waves above the plasma frequency.

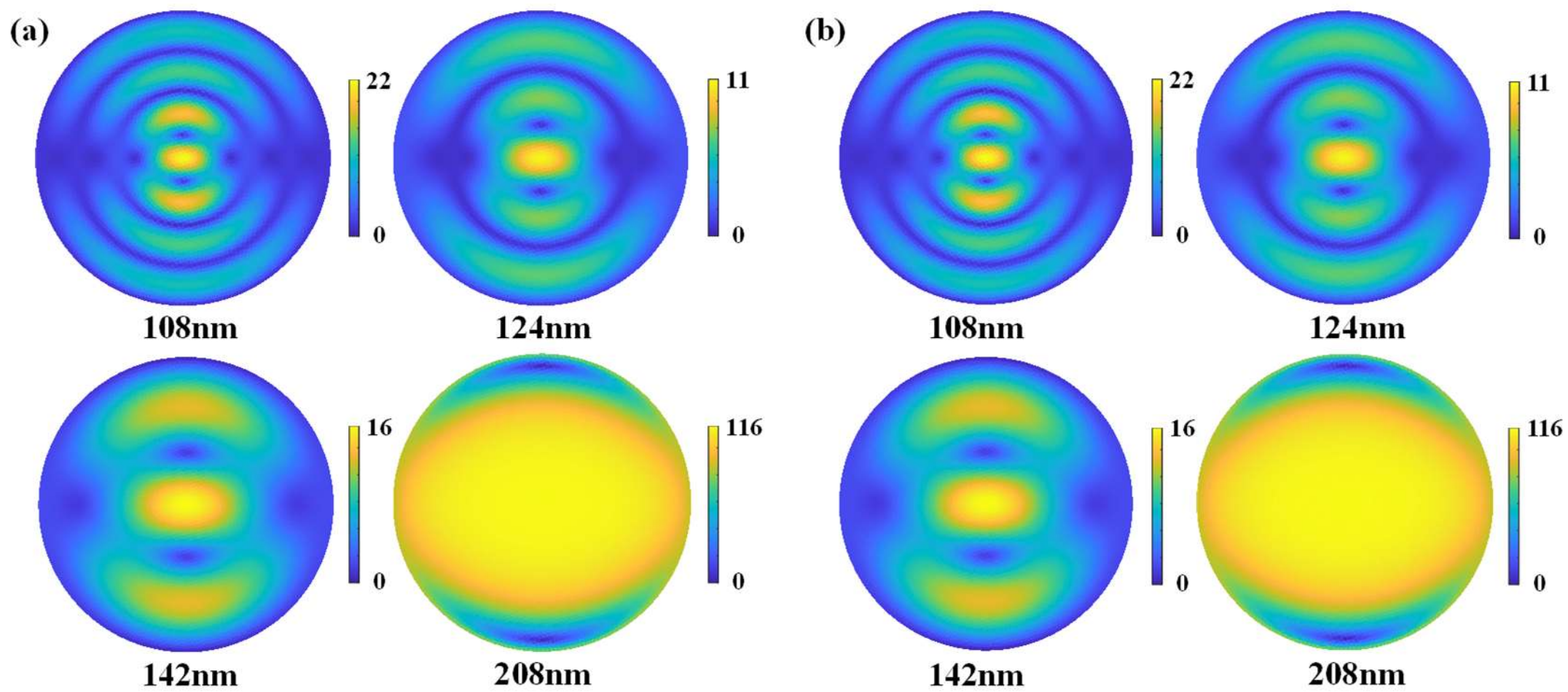

Fig. 4. Illustration of the electric field within the wire on the $xy$ plane. (a) is obtained from the proposed algorithm, while (b) is from the in-house BEM. The four wavelengths correspond to the four resonances of the nonlocal case shown in Fig. 3(a). In this figure, the colors are coded from blue to yellow to denote the magnitude of the fields. The same color bar will be used for the rest of the text.

Further, at the resonances we collect the electric fields in the nanowire on the $xy$ plane with the surface area of the wire being discretized by 27711 triangles. Then, the results are compared with simulations using the in-house BEM solver [17], shown in Fig. 4, where a good agreement can be observed. The meshes used for calculation are same for the two methods, where the max area of the triangles is 0.000456 $nm^2$ and the max radius of their inscribed circles is 0.00926 nm, and the electric fields are evaluated at the centroids of all triangles. The resonant modes are at 108 nm, at 124 nm, at 142 nm, and at 208 nm, respectively. For the mode at 108 nm, the mean absolute relative error is 0.9253%, and the mean squared relative error is 0.0020%; for the mode at 124 nm, the mean absolute relative error is 0.3750 %, and the mean squared relative error is 0.3142%; for the mode at 142 nm, the mean absolute relative error is 0.2499 %, and the mean squared relative error is 0.00081%; for the mode at 208 nm, the mean absolute relative error is 1.0273 %, and the mean squared relative error is 0.00014%.

For the sake of completeness, the runtime of both solvers without postprocessing is compared. It takes BEM 29.84 s to finish, while for the proposed solver, it is 1.47 s. As a comment, in our calculation, the proposed method uses $m$ up to 5 and the BEM solver uses 3483 line segments to discretize the cylindrical IF on the $xy$ plane.

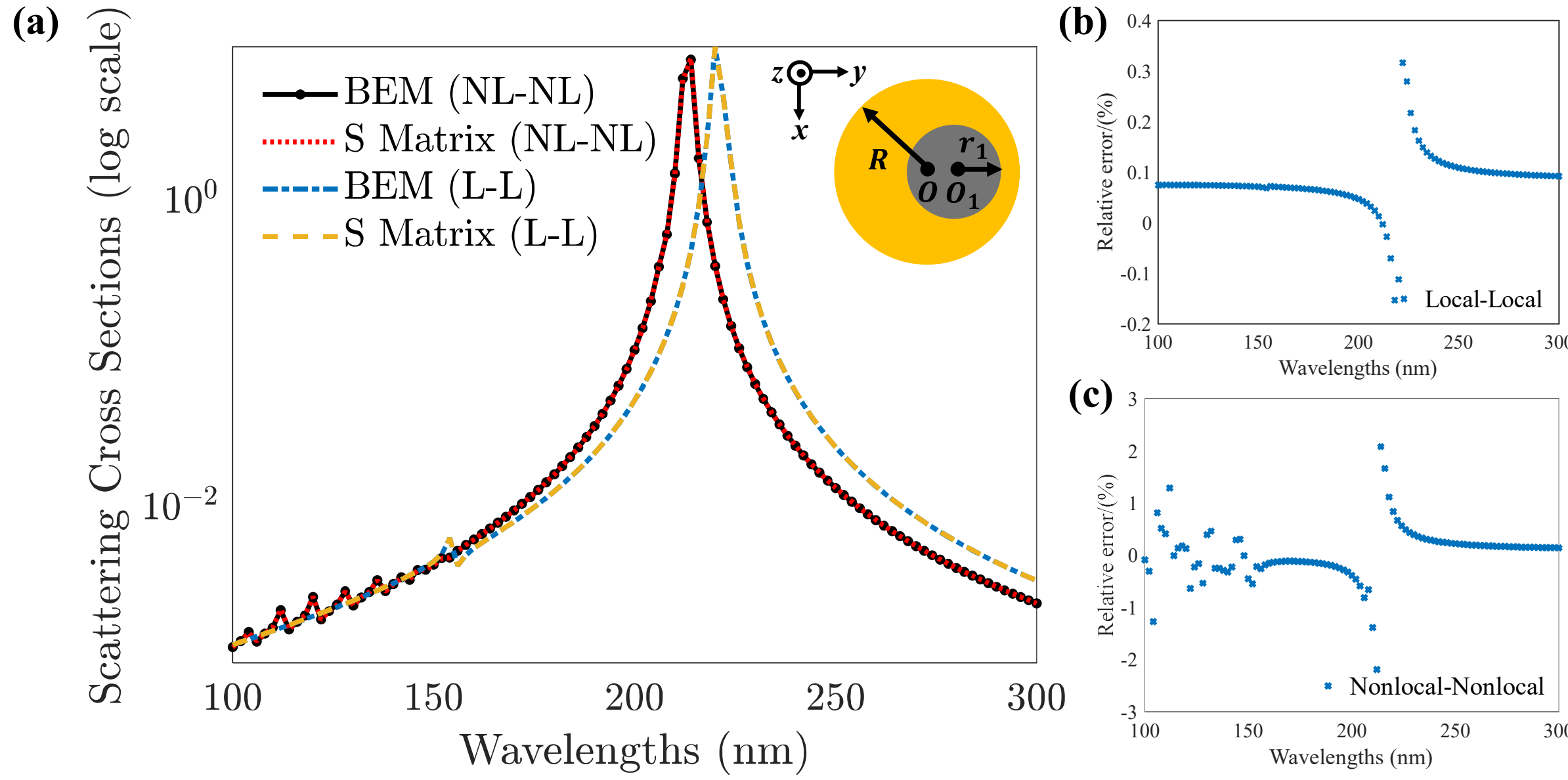

Fig. 5. Simulation results for an eccentric nanowire. (a) Evaluated scattering cross sections in a log scale. The inset shows the cross section of an eccentric dual-material nanowire in the $xy$ plane. The material is treated as a local medium by using the BEM (black dotted line) and proposed S Matrix Method (red dashed line); and as a nonlocal medium by using the BEM (blue line) and the proposed S Matrix Method (yellow dashed line). In (b) and (c), the relative errors as defined in (48) are shown.

*B. Eccentric Dual-material Nanowire*

We consider an eccentric dual-material nanowire (see the inset in Fig. 5(a)). The structure is placed in vacuum and has two cylindrical IFs. The outer IF separates gold (Au, the inner region) from vacuum (the outer region). The center of the outer IF is set as the origin $O$ of the coordinate system and its radius is $R = 2$ nm. The inner IF separates silver (Ag, the inner region) from gold (Au, the outer region) and is centered at $o_1$ $(0, 0.3, 0)$ with a radius $r_1$ of 1 nm. Here, we consider two cases: 1) both Au and Ag are treated as local media (LRA); 2) both Au and Ag are treated as nonlocal media (HDM). For the HDM, the plasma frequency and the damping rate of Au is given above. For Ag, the Drude damping rate is $7.29 \times 10^{13}$ rad/s, taken from tabulated data [75], and the Fermi velocity of Ag is $1.39 \times 10^6$ m/s [76]. The effective plasma wavelength of Ag is approximately 149.7 nm, where the oscillator strength of the free-electron Drude contribution $f_0$ is 0.845. The structure is excited by a TE plane wave travelling along the positive $y$ direction and polarized along the $x$ axis. The amplitude of the incident electric field is 1 V/m. Again, we consider a wavelength range from 100 nm to 300 nm with 101 sampling points taken in between.

As the result of the simulations, the scattering cross sections $\sigma_{sca}$ are evaluated in the same way as in the first example. The results are also compared with those predicted by the in-house BEM solver [17]. In Fig. 5, the scattering cross sections show good agreement, with the maximum relative error being within 3%. Like the previous example, the largest errors appear at the resonant wavelengths. Further, the near-field mappings of resonant modes are also calculated (see Fig. 6) and compared with the results obtained by BEM. The electric fields are collected over the surface area within the core and are evaluated at the centroids of all discretized triangles. The core is discretized by 2356 triangles, and the maximum area of the triangles is $0.0017843$ nm$^2$, while the maximum radius of their inscribed circles is $0.018284$ nm. The shell is discretized by 7013 triangles, and the maximum area of the triangles is $0.001993$nm$^2$, while the maximum radius of their inscribed circles is $0.019529$ nm.

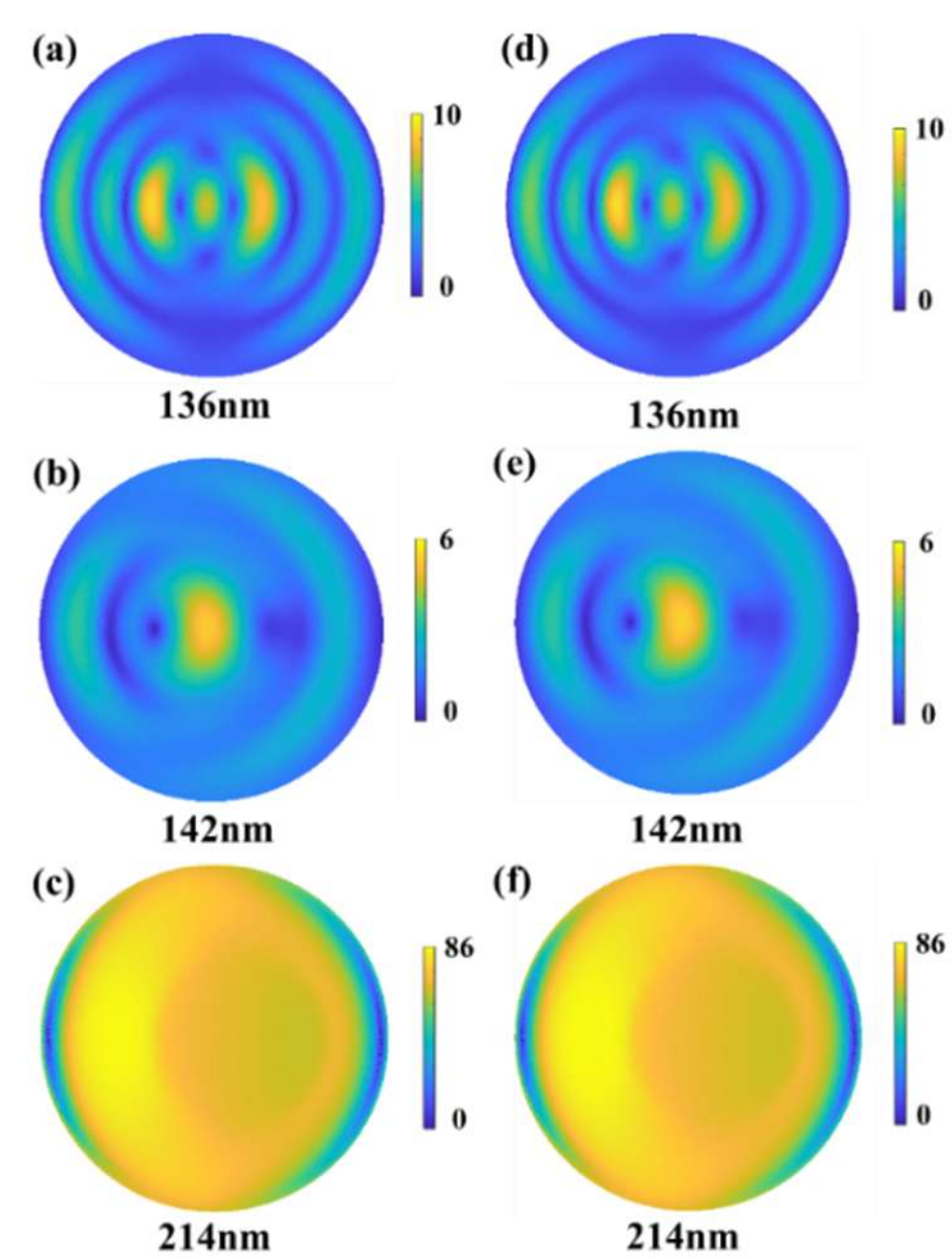

Fig. 6. Illustration of the electric field distribution within the eccentric nanowire on the $xy$ plane obtained from the proposed algorithm and the in-house BEM code. The distributions are plotted for the resonant wavelengths in Fig. 5(a). (a-c) Near-field mapping obtained by the proposed **S**-Matrix Method. (d-f) Near-field mapping obtained by the BEM.

For the mode at 136 nm, the mean absolute relative error is 4.1454%, and the mean squared relative error is 0.3142%; for the mode at 142 nm, the mean absolute relative error is 2.7588%, and the mean squared relative error is 0.1621%; for the mode at 214 nm, the mean absolute relative error is 1.0273%, and the mean squared error is 0.0107%. As for the runtime of both solvers without postprocessing is compared. It takes BEM 495.30 s to finish, while for the proposed solver, it is 2.48 s. The proposed method uses $m$ up to 10 and the BEM solver uses 13960 line segments to discretize the cylindrical IF on the $xy$ plane.

*C. Self-similar Sodium Nanolens*

We compare the optical response of the LRA, the HDM, and the SRM from a nanolens with decreasing radii and gaps inspired by [77] (see Fig. 7). The structure is placed in vacuum and has three IFs. The IFs are named IF1, IF2 and IF3, and each IF separates sodium (Na, the inner region) from vacuum (the outer region). The origin $O$ of the coordinate system is set at the center of IF2 (see Fig. 7), and the radii of the IFs are $R_1 = 18$ nm, $R_2 = 6$ nm, and $R_3 = 2$ nm. The gap $g_1$ between IF1 and IF2 is 3nm, while the gap $g_2$ between IF2 and IF3 is 1 nm. The gap size decreases in accordance with the ratio of IF's radii, which is 3 in this case. Here, we use $m$ up to 20. For the HDM, the plasma wavelength is approximately 210.45 nm and the Drude damping rate of Na is $1.5193 \times 10^{14}$ rad/s, taken from data in [21]. The Fermi velocity of Na is $1.06 \times 10^6$ m/s [76]. The structure is excited by a plane wave travelling along the $y$ direction and polarized along the $x$ axis. The amplitude of the incident electric field is as before 1 V/m. Here, we consider a wavelength range from 200 nm to 500 nm within 301 wavelength points in between (shown in Fig. 8).

At the targeted main resonance (where the field is focused at the smallest gap and thus enhanced), we plot the field on a cut along the $xy$ plane. The size of the cut is 64 nm by 40 nm and the cut is discretized by 231212 triangles. The maximum area of the triangles is $0.030204\ \text{nm}^2$ and the maximum radius of their inscribed circles is $0.075497$ nm. The simulations are conducted in the proposed algorithm. We consider three cases: 1) Na is treated as a local medium; 2) Na is treated as HDM metal; and 3) Na is treated by the SRM.

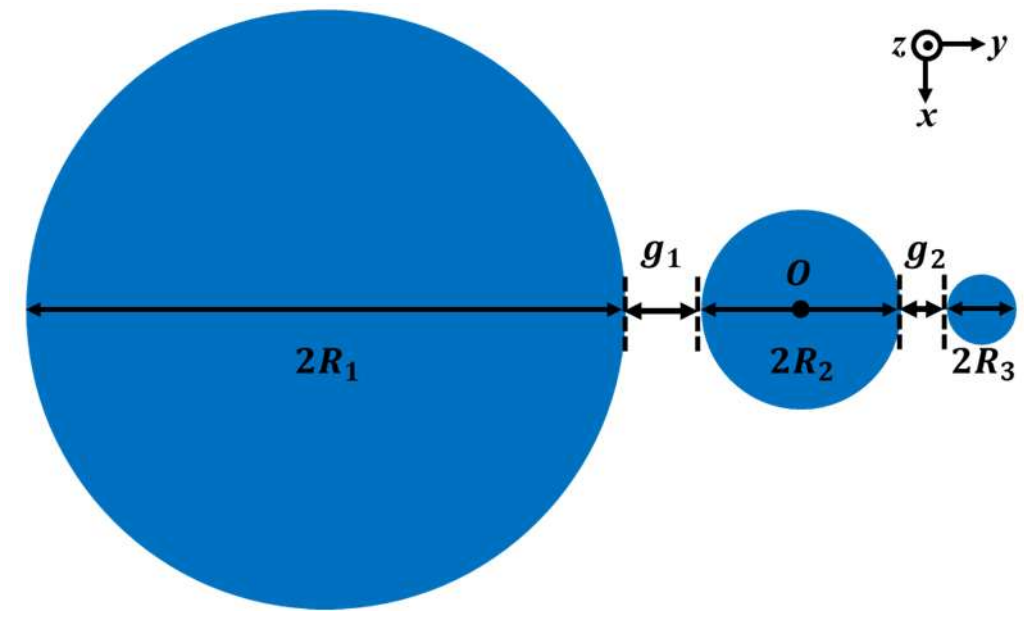

Fig. 7. Illustration of a plasmonic self-similar cylindrical nanolens and there are three IFs (IF1, IF2 and IF3 from the left to the right) in total. The origin $O$ of the coordinate system is set at the center of the second IF. The radii of the IFs are $R_1 = 18$ nm, $R_2 = 6$ nm, and $R_3 = 2$ nm. The gap $g_1$ between IF1 and IF2 is 3nm, while the gap $g_2$ between IF2 and IF3 is 1nm.

In the higher frequency region, the two hybridized collective resonances observed within the LRA are no longer separately resolved for the HDM and SRM. For the HDM, the size-dependent nonlocal blueshifts modify the spectral matching and hybridization among the individual cylinder resonances, while surface-induced damping introduced by the SRM additionally broadens the overall spectral response. Further details are provided in Appendix A. At the targeted main resonance, the optical response predicted by the HDM shows a blue-shift, while that of the SRM shows a red-shift, compared to the main resonance obtained by the LRA (see the peak around 374 nm of the red-dotted line and the peak around 409 nm of the yellow dashed line). To explain, the HDM pushes the induced charges inside the metal of the metal-dielectric IF, suggesting that the total restoring force in plasmonic oscillations is increased, causing a blueshift of the Localized Surface Plasmons (LSPs). In contrast, owing to the electron spill-out and the relaxation of the hard-wall boundary condition, the resonances predicted by implementing SRM are redshifted. The spectral broadening is related to Landau damping, a nonlocal dissipative mechanism, causing strong decrease in the amplitude of the scattering cross sections. The electric field mapping is also depicted for the LRA, the HDM and the SRM, of which the field enhancement at the smallest gap happens at 386 nm, 374 nm, and 409 nm. For the LRA, the field enhancement is 154; for the HDM, 93; and for the SRM, 130. The convergence test on the maximum field enhancement regarding $M$ is given in Appendix B. Such a reduction is because both the HDM and the SRM remove the surface charge accumulation at the boundaries. On the other hand, the "spill-in" effect in the HDM decreases the electric size of the plasmonic nanowire, while the "spill-out" effect in SRM increases the electric size of the structure, making the detection points "further away" or "closer to" the IF, leading to smaller or larger field enhancement.

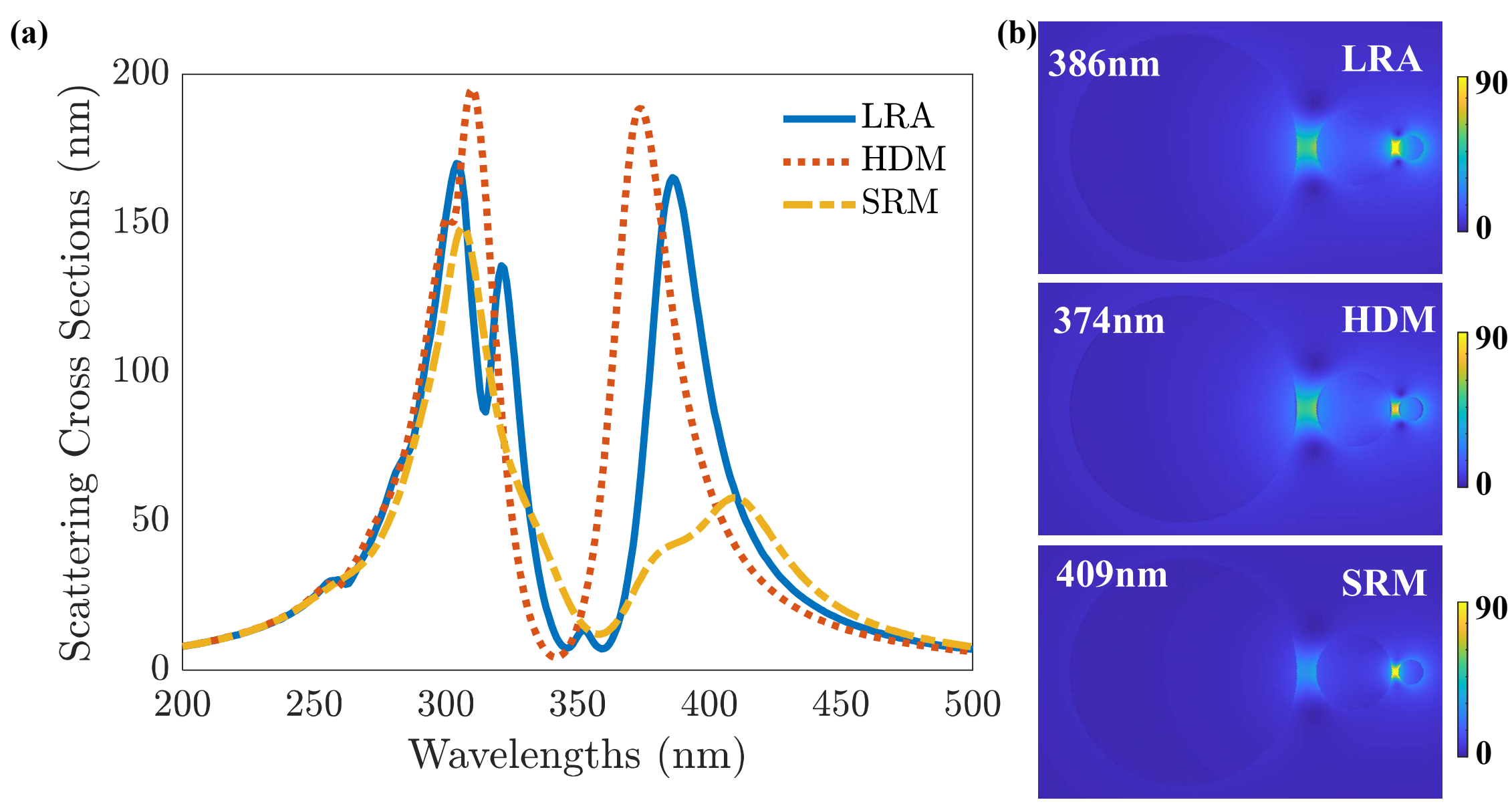

Fig. 8. Optical response from Na nanolens. (a) Scattering cross sections obtained by 1) treating Na as a local medium (LRA, blue solid line); 2) treating Na as a nonlocal medium (HDM, red dotted line); and 3) using the SRM (yellow dashed line). (b) The electric field (magnitude) distributions along a cut made on the $xy$ plane at 386nm, 374nm, and 409nm for the three different models.

## V. CONCLUSION

In conclusion, we propose an effective and efficient algorithm to predict the nonclassical optical response from a system of metallic nanocylinder aggregates. The algorithm can properly account for two important semiclassical material models, that is, the HDM and the SRM. This requires a systematic adaptation of the S-matrix and the translational addition theorem in a unified scheme. The proposed method is validated against an in-house developed BEM solver. Good agreements are seen in both the scattering spectra and the near field, with the resonant modes due to longitudinal waves being clearly observed. Different optical responses predicted by the LRA, the HDM and the SRM are further compared for a self-similar sodium nanolens, in accordance with previously reported physical discoveries. The proposed method provides a generalized framework for exploring nonclassical optical responses in systems of multiple plasmonic nanowires and holds promise as a valuable tool for studying mesoscopic electrodynamics in applications such as plasmonic waveguides and optical fibers.

## APPENDIX

### *A. Analysis of the Nanolens Resonances*

Here, we discuss the resonances of the nanolens in the higher frequency range from 280 nm to 340 nm, as shown in Fig. 8. For the nanolens, the LRA resonances occur at 304 nm and 320 nm, while the HDM and SRM resonances occur at 309 nm and 306 nm, respectively. Fig. A1 shows the scattering cross sections of the individual cylinders. For the LRA, the resonances of cylinders 1, 2, and 3 appear at 312 nm, 301 nm and 298 nm, respectively. The corresponding HDM resonances are located at 311 nm, 297 nm and 288 nm, respectively, showing an obvious size-dependent blueshift. For the SRM, the resonances occur at 313 nm, 305 nm and 311 nm, respectively, demonstrating a clear redshift relative to the LRA.

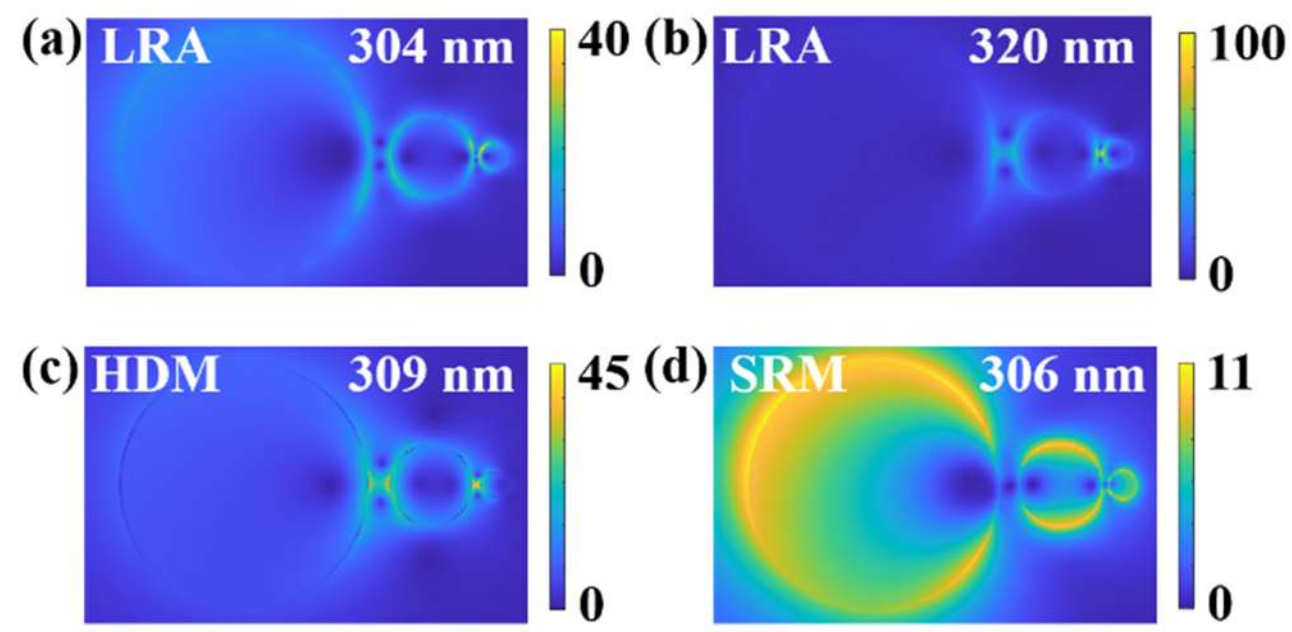

Fig. A2. The electric field (magnitude) distributions along a cut made on the *xy* plane at the 304 nm, 320 nm for the LRA, 309 nm for the HDM, and 306 nm for the SRM.

The two LRA resonances at 304 and 320 nm exhibit distinct field distributions, supporting their interpretation as different hybridized collective responses, as shown in Fig. A2. The 304 nm resonance exhibits a relatively distributed field profile, while for the 320 nm resonance, field is more localized in the nanogaps. Within the HDM, the size-dependent nonlocal blueshift modifies the resonance wavelengths, and, consequently, the spectral matching and mode hybridization among the cylinders. The resulting spectral detuning may weaken and broaden the collective response associated with the 304 nm LRA peak, possibly accounting for the small shoulder in the HDM spectrum. Based on the similarity of the field distributions, the resonance at 309 nm might correspond to the blue shifted counterpart of the LRA resonance at 320 nm. Within the SRM, the surface induced damping and spectral broadening suppress the weaker hybridized contributions and broaden the overall spectral response. The resonance at 306 nm exhibits a markedly different and much more spatially distributed field profile.

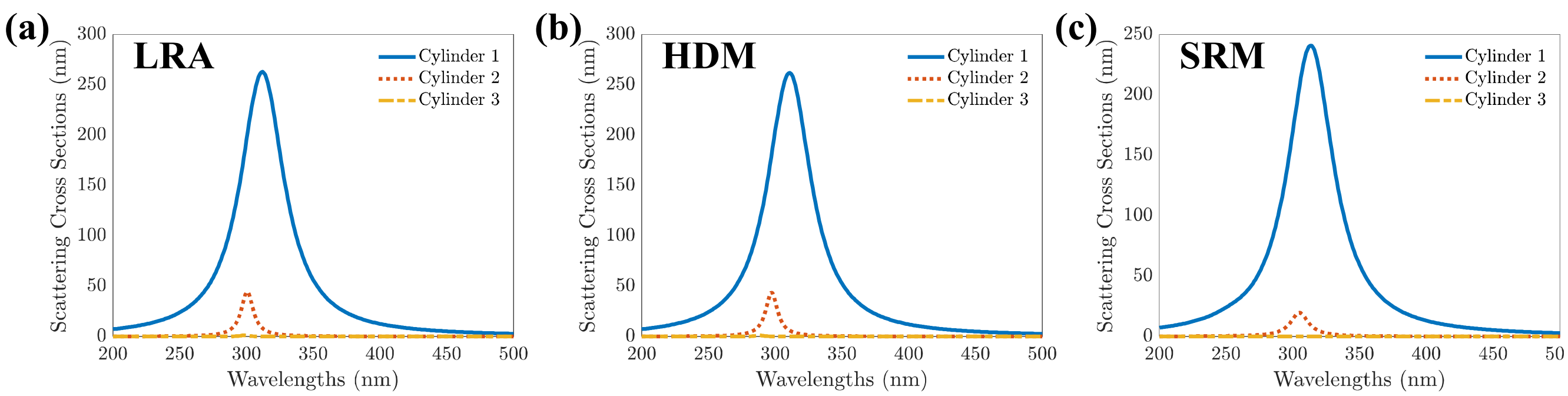

Fig. A1. Optical responses of the three individual cylinders in the nanolens. (a) Scattering cross sections obtained within the LRA. (b) Scattering cross sections obtained within the HDM. (c) Scattering cross sections obtained within the SRM.

### B. The convergence test

Here, a near-field convergence test with respect to different $M$ is performed while the meshes used for discretizing stay unchanged. For all $M$'s and each model, the maximum field was consistently observed at the same spatial position $(6.9756, 0.10505, 0)$ nm.

| | $M = 10$ | $M = 12$ | $M = 15$ | $M = 18$ | $M = 20$ |
|---|---|---|---|---|---|
| **LRA** | 149.3698 | 152.6153 | 154.3450 | 154.8146 | 154.9152 |
| **HDM** | 92.2940 | 92.8918 | 93.1469 | 93.2003 | 93.2095 |
| **SRM** | 132.9143 | 132.2342 | 131.8107 | 130.7117 | 130.1669 |

Table A1. The maximum field enhancement obtained under three physical models with different cutoff parameters.

As shown in Table A1, when $M$ increases from 18 to 20, the maximum field enhancement changes by only $0.065\%$, $0.010\ \%$, and $0.417\ \%$ for the LRA, HDM, and SRM, respectively. These sub-percent relative differences provide numerical evidence of satisfactory near-field convergence, although the simulation of the SRM converges more slowly than the other two models. In the manuscript, $M = 20$ is adopted and values of maximum field enhancement are given with appropriately limited numerical precision.